\documentclass[]{nst}

\usepackage{subfigure,dcolumn}
\usepackage[T2A,T1]{fontenc}
\usepackage[russian,english]{babel}
\usepackage{multirow} 
\usepackage{listings}
\begin{document}

\title{Neural network study on nuclear ground-state spin distribution within random interaction ensemble}\thanks{Supported by the National Natural Science Foundation of China Youth Fund (12105234)}

\author{Deng Liu}
\affiliation{School of Mathematics and Physics, Southwest University of Science And Technology, Mianyang, 621010, China}

\author{Alam Noor A}
\affiliation{School of Mathematics and Physics, Southwest University of Science And Technology, Mianyang, 621010, China}

\author{Zhenzhen Qin}
\email[Corresponding author, ]{qin_zhenzhen@hotmail.com.}
\affiliation{School of Mathematics and Physics, Southwest University of Science And Technology, Mianyang, 621010, China}

\author{Yang Lei}
\affiliation{School of  Nuclear Science and Technology, Southwest University of Science And Technology, Mianyang, 621010, China}

\begin{abstract}
The distribution of nuclear ground-state spin in the two-body random ensemble (TBRE) is studied by using a general classification neural network (NN) model with the two-body interaction matrix elements as input features and corresponding ground-state spins as labels or output predictions. It seems that quantum many-body system problem exceeds the capability of our optimized neural networks when it comes to accurately predicting the ground-state spin of each sample within the TBRE. However, our neural network model effectively captures the statistical properties of the ground-state spin. This may be attributed to the fact that the neural network (NN) model has learned the empirical regularity of the ground-state spin distribution in TBRE, as discovered by human physicists.
\end{abstract}

\keywords{neural network; two-body random ensemble; spin distribution of nuclear ground state.}

\maketitle

\section{Introduction}

The atomic nucleus is a typical complex many-body quantum system. Conventionally, one needs to construct the many-body Hamiltonian or Lagrangian based on reliable interactions in order to investigate this complex system. However, such a task is usually challenging, as in many-body problems interactions are strongly entangled with structures, and thus the self-consistent requirement under a certain ansatz leads to a vague, or at some degree inaccurate, many-body Hamiltonian. Fortunately, if one is only interested in regularity and robust properties of many-body system that are independent of interaction details, the vagueness of Hamiltonian provides an alternative perspective, with random number as some parameters of nuclear interactions, i.e., random interactions, to statistically probe those robust regularity of nuclei.

The study of random interactions can be traced back to the investigation of Wigner's random matrices theory (RMT) \cite{1}, where random numbers were used as matrix elements of the many-body Hamiltonian. By diagonalizing these random matrices, one can obtain spectral statistical properties that agree with experimental data. The spectral properties of RMT were further linked to quantum chaos \cite{2}. In the 1970s, Wong, Bohigas, $et ~al.$ \cite{3, 4, 5} introduced the idea of randomizing two-body interaction matrix elements in shell-model calculations \cite{6, 7} to quantitatively demonstrate the phenomenon of quantum chaos in nuclei \cite{5, 8, 9, 10,11}. The shell-model calculations with random interactions create an ensemble of virtual nuclei. Such an ensemble is known as the two-body random ensemble (TBRE). The study with TBRE has revealed that certain robust features of nuclei do not necessarily depend on the specific details of the interaction.

Following this philosophy, Johnson, Bertsch, $et~al.$ \cite{12, 13} reported a series of robust and interaction-independent statistical properties of low-lying states in nuclei. One of the most notable findings is the "predominance of spin-zero ground state" in even-even nuclei. Even-even nuclei exhibit a considerably higher probability of having spin-zero ground states compared to the fraction of zero-spin configurations in the entire shell-model space. Later on, such a phenomenon was also observed in the Interacting Boson Model (IBM) \cite{14, 15,16}. The spin-zero ground states of even-even nuclei are conventionally attributed to the short-range nature of the nuclear force. However, in the TBRE, interactions are entirely random, and no specific force predominates. The predominance of the spin-zero ground state of TBRE somehow contradicts the conventional understanding of how spin-zero ground states emerge from even-even systems. Therefore, many efforts have been devoted to understanding this robust property of the TBRE, which has proven to be significantly challenging and reflects the complexity of the quantum many-body problem. Some phenomenological attempts include the studies of the distribution of the lowest eigenvalues for each spin \cite{14} and its width \cite{17}, the geometric chaos of spin coupling \cite{18}, the maximum and minimum diagonal matrix elements \cite{19}, the IBM-limit of spin distribution in the IBM with TBRE \cite{20, 21, 22}, the wave-function properties of different spin ground states \cite{23, 24}, energy scale features of different spin ground states \cite{25}, and the correlation between the probability of zero-spin ground states and the central values of the distribution of two-body matrix elements \cite{26}. To explain this phenomenon, it is necessary to mathematically calculate the probability distributions of various spin states as ground states. However, nuclear models are typically nonlinear systems that are difficult to apply statistical theories to. Therefore, several empirical rules have been proposed to predict the probability distribution of ground state spins. For example, Kusnezov $et~al.$ used the random polynomial method \cite{24} to a priori determine the probability distribution for $sp$ bosons, which yielded consistent results with those obtained by Bijker $et~al.$ using mean-field methods \cite{21, 22}. Chau and others discussed the cases of $d$ boson systems and four fermions in the $f_{7/2}$ shell, demonstrating the correlation between specific ground states and the geometric shapes determined by nuclear observables and predicting the probabilities for the ground-state spin \cite{27}. { Zhao $et~al.$ suggested that the spins of ground states in the TBRE may be associated with specific two-body interaction matrix elements, and thus proposed an empirical approach \cite{28} to predict the distribution of ground-state spins. The correlation between the ground state spin and the two-body interaction matrix elements in this empirical approach is also crucial in our work.}

Since the non-linearity of nuclear model is way complex to overcome, one can take a bypass to touch the origin of the predominance of zero-spin ground states, by using a non-linear but simple enough model to simulate the behavior of the shell model, and studying the spin determination mechanism therein, which may provide more insight from a different prospective. The neural network model (NN) can be potential candidate for such simulations with its powerful learning, prediction, and adaptation capabilities, which have been successfully applied in diverse fields such as language translation, speech recognition, computer vision, and even complex physical systems \cite{29, 30, 31, 32}. More specifically, the NN models have been extensively utilized in nuclear structure studies to predict various unknown nuclear properties using existing experimental data. These properties include mass \cite{33, 34, 35}, charge radii \cite{36,37}, low-lying excitation spectra \cite{38, 39}, $\beta$ decay lifetimes \cite{40}. However, most of these works only made best use of the fitting capacity of the NN, without fully exploring its classification capability for nuclear structure research.

In this work, we make a tempt to distinguish samples with different ground-state spin in the TBRE, by adopting the classification capability of the NN with supervised learning. The adopted neural network (NN) is trained using the interaction matrix elements from TBRE samples as features and the ground state spin as the label. { In this process, the NN learns the behavior of the ground state spin in TBRE, as well as the specific correlations between interaction elements and the ground state spin, as described in the empirical approach \cite{28}.} A notable advantage of using NN in the TBRE study lies in the ability of the TBRE to provide nearly infinite independent samples for the NN training, which avoids the over-fitting, and thus potentially enhance the generalization ability of the NN and facilitating the simulation of the shell model production of the ground-state spin. We will fully present the performance of the NN in predicting the ground-state spins, and reproducing their distribution in the TBRE. The neural network architecture proposed in this paper may serve as a valuable benchmark for other classification-based applications.

\section{MODEL FRAMEWORK}
\subsection{Two-Body Random Ensemble (TBRE)}

In the TBRE, the nuclear Hamiltonian only includes two-body interactions expressed as follows:
\begin{equation}\label{eq1}
		H=\sum_{J} \sum_{j_{1} j_{2}} \sum_{j_{3} j_{4}} G^{J}_{j_{1} j_{2} ; j_{3} j_{4}} A_{J }^{\dagger}\left(j_{1} j_{2}\right) A_{J }\left(j_{3} j_{4}\right).
\end{equation}
In Eq. (\ref{eq1}), $G^{J}_{j_{1} j_{2} ; j_{3} j_{4}}$ represents the matrix elements of the two-body interaction, $A_{J }^{\dagger}\left(j_{1} j_{2}\right)$ denotes the creation operator of the nucleon pair with two nucleons on the $j_1$ and $j_2$ orbits coupled to total angular momentum $J$, and similarly, $A_{J }\left(j_{3} j_{4}\right)$ corresponds to the annihilation operator of nucleon pair.

In TBRE, the matrix elements $G^{J}_{j_{1} j_{2} ; j_{3} j_{4}}$ in Eq. (\ref{eq1}) are independent random numbers following the Gaussian distribution with probability function:
\begin{equation}\label{eq2}
	f\left(G^{J }_{j_{1} j_{2} j_{3} j_{4}}\right)=\frac{1}{\sqrt{2 \pi} \sigma} \exp\left\{-\frac{\left(G^{J }_{j_{1} j_{2}; j_{3} j_{4}}\right)^{2}}{2 \sigma^{2}}\right\} \text { , }
\end{equation}
where
\begin{equation}
	\sigma^{2}=\frac{1}{2}\left(1+\delta_{j_{1} j_{3}}\delta_{ j_{2} j_{4}}\right) \text { , }
\end{equation}
to maintain the statistical distribution of the interaction matrix elements invariant during the arbitrary single-particle transformation.

\subsection{Classification neural network}

\begin{figure}[!htb]
	\includegraphics
	[width=1.0\hsize]
	{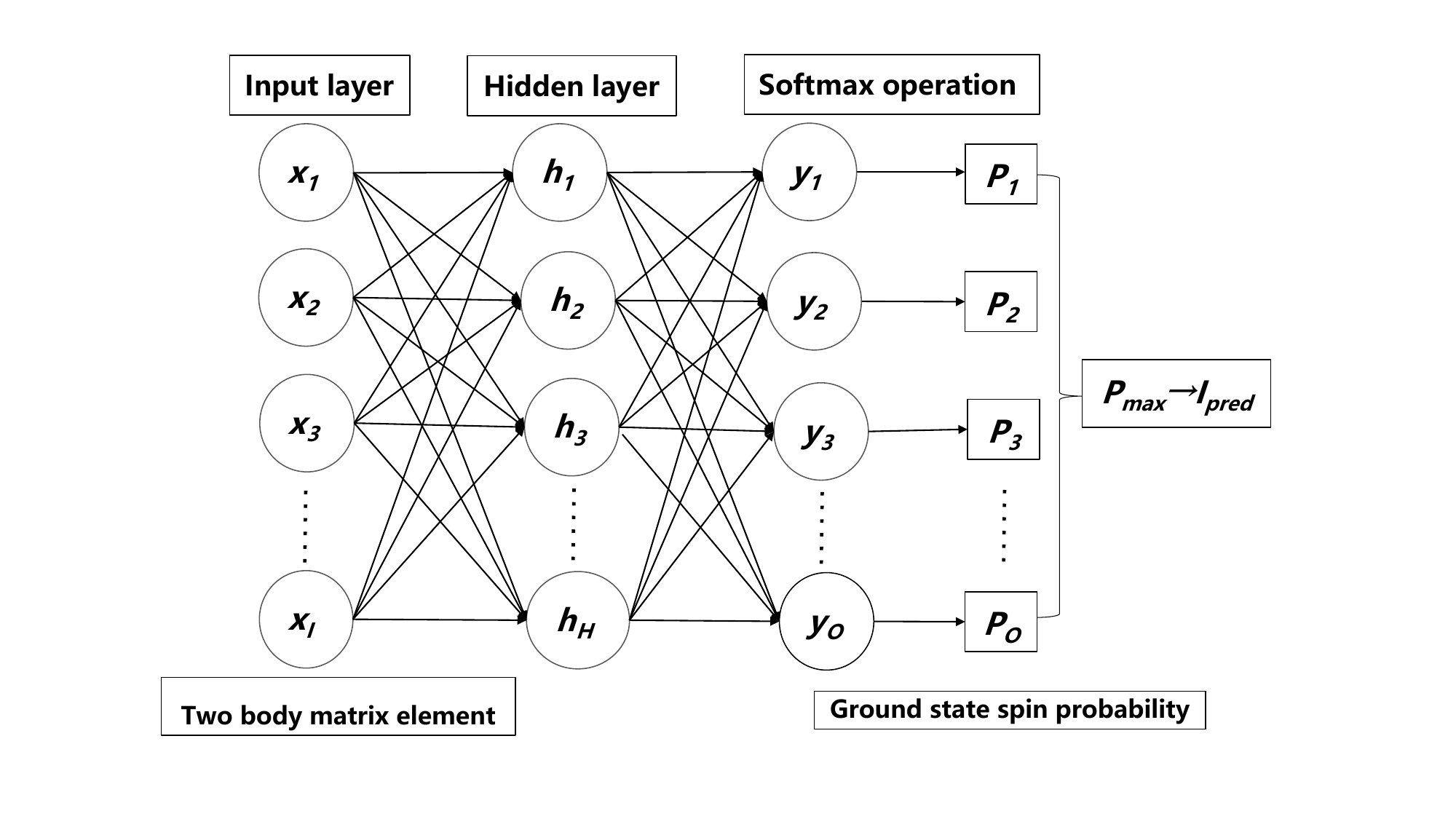}
	\caption{Schematic diagram of the adopted neural network classification model.}
	\label{fig:1}
\end{figure}

The classification model in this paper utilizes a neural network, which consists of an input layer, one or more hidden layers, and an output layer. This structure is illustrated in Fig. \ref{fig:1}, (with one hidden layer shown as an example). The input layer receives the matrix elements of the two-body interactions in the shell model, specifically the $G^{J}_{j_{1} j_{2} ; j_{3} j_{4}}$ values in Eq. (\ref{eq1}), with the number of inputs equal to the number of independent two-body interaction matrix elements in a specific shell-model space. The output layer provides the probabilities of different spin states being the ground state based on the corresponding input interactions. The number of outputs should be equal to the number of possible ground-state spins.
The activation function used in this model is the Rectified Linear Unit (ReLU) function \cite{41}, which will be justified with Table \ref{tab:2} later. Assuming vector $\vec x=\left\{x_{i}\right\}$ represents the network input, i.e., the $G$ two-body interaction matrix elements in Eq. (\ref{eq1}), and $\vec y$ is the network output, whose elements correspond to the probability of each spin being the ground-state spin. The relationship (with one hidden layer) can be expressed analytically as follows:
\begin{equation}\label{eq4}
	y_k(\vec x;\vec\omega)=a_k+\sum_j b_{kj} {\rm ReLU}  \left(c_{j}+\sum d_{j i} x_{i}\right), 
\end{equation}
where $\vec \omega=\{a_k, b_{kj}, c_{j}, d_{j i}\}$ represents the parameter vector of the neural network.

The output layer introduces the Softmax function\cite{42}, which transforms the unnormalized output values into non-negative probability values that sum up to 1.
\begin{equation}\label{eq:pk}
	P_k=\left.Softmax(\vec y)\right|_k=\frac{e^{y_{k}}}{\sum_{k} e^{y_{k}}}.
\end{equation}
This operation preserves the differentiability property of the model, as well as the relative order of unnormalized output values. It also allows the model's output to be interpreted as probabilities for each class, facilitating the direct interpretation and utilization of these probabilities for classification decisions. Therefore, it is frequently employed in neural network models for classification problems. Here, $P_k$ is the probability of $k$-th spin to be the ground-state spin. Thus, the maximum of $P_k$ determines the ground-state spin according to the $\vec x$ feature, i.e., the inputted two-body matrix elements. All the elements $P_k$ construct the predicted probability $\vec P$ vector from the neural network model.

To train the NN model, firstly, we prepare a training set consisting of $N$ samples, $D=\left\{\left(\vec x_1,S_1\right)\right.,\left.\left(\vec x_2,S_2\right),\ldots,\left(\vec x_N, S_N\right)\right\}$ out of $\sim$100,000 shell-model calculations, where $\vec x_i$ includes two-body interaction matrix elements in a single shell-model calculation, and $S_i$ is corresponding ground-state spin from such a shell-model. Secondly, for each $S_i$ spin, we create the label $\hat {\vec P}^i$ vectors, which is a hot-one vector, and only include one non-zero elements of value ``1'', corresponding to a 100\% probability of $S_i$ ground-state spin, and 0\% probabilities of the rest other spins. Thirdly, we define the loss function to evaluate the similarity between the label $\hat {\vec P}^i$ vector and the NN predicted $\vec P^i$ vector from Eq. (\ref{eq:pk}) as
\begin{equation}\label{eq5}
	loss(\vec P^i,\hat {\vec{P}}^i)=-\sum P^i_{m} \log \hat{P}^i_{m},
\end{equation}
which is the common the loss function for training the NN model for classification problems. With the training samples, and corresponding loss function, we train our network by adjusting the network parameter vector $\vec \omega$ with Adam (Adaptive Moment Estimation) optimization algorithm \cite{43} to minimized the sum of the loss functions for all training samples. Consequently, a neural network model with predictive capabilities is obtained.

\subsection{Shell Model Spaces}\label{sec:sm}

We perform approximately 100,000 TBRE calculations in six model spaces. These include four valence nucleons in the $f_{7/2}$ orbital virtual nuclear (Simply expressed as $(f_{7/2})^4 $), four valence nucleons in the $ h_{11/2} $ orbital virtual nuclear (Simply expressed as $ (h_{11/2})^4 $), and two, four, and six valence neutrons in the $ sd $ shell (corresponding to $ ^{18} $, $ ^{20} $, and $ ^{22} $ Ne nuclear, respectively), and six valence neutrons in the $ pf $ shell (corresponding to the $^ {46} $Ca nuclear). These six model spaces represent various levels of many-body complexities.

In the $(f_{7/2})^4$ space, the eigenvalues from the shell model are simple linear combinations of two-body interaction matrix elements, as demonstrated in Eq. (1.81) in reference \cite{44}. The ground state spin corresponds to the lowest eigenvalue associated with that specific spin. A neural network without a hidden layer corresponds to linear combinations of input two-body interaction matrix elements, followed by the application of the softmax operation to identify the smallest linear combinations. The calculation process of the ground state spin determination is somewhat similar in both models, where the weight parameters, i.e., $d_{ji}$ parameters in Eq. (\ref{eq4}), in the neural network correspond to the cfp coefficients \cite{44} in the shell model, and the softmax input to the neural network is equivalent to the energy eigenvalues of the shell model. Since a hidden layer not only complicates the neural network (NN) model but also violates the correspondence between the NN and the shell model, we chose to exclude the hidden layer in our neural network model for the $(f_{7/2})^4$ space.

Regarding the $(h_{11/2})^4$, $^{18}$Ne space, the complexity increases beyond the $(f_{7/2})^4$ space. Some eigenvalues are still a linear combination of two-body interaction matrix elements, while others have to be obtained through diagonalization. Although some diagonalization processes with dimensions less than 5 are analytical, it is no longer possible to relate the weight parameters $d_{ji}$ to cfp coefficients in those processes. Therefore, the hidden layers can enhance the adaptability of neural network models to nonlinear diagonalization \cite{45}. In the $^{20,~22}$Ne and $^{46}$Ca space, the relationship between the eigenvalues and cfp coefficients has become completely nonlinear. Some of these relationships are transcendental, which makes hidden layers even more necessary. Table \ref{tab:1} presents the TBRE sample sizes and input-output settings of our neural network models for different spaces.
\begin{table}[!htb]
	\caption{Input-output settings for the six model spaces. The input number corresponds to the number of two-body interaction matrix elements, while the output number corresponds to the number of possible ground state spins.}
	\label{tab:1}
	\begin{tabular*}{8cm} {@{\extracolsep{\fill} } ccc}
		\toprule
		Model Space &Input & Output \\ 
		\midrule
		$(f_{7/2})^4$  & 4  & 5  \\ 
		$(h_{11/2})^4$ & 6  & 10 \\
		$^{18}$Ne     &  30 & 5  \\
		$^{20}$Ne     &  30 & 7  \\
		$^{22}$Ne     &  30 & 8   \\
		$^{46}$Ca     &  94 & 13 \\ 
		\bottomrule
	\end{tabular*}
\end{table}

\subsection{Optimization of network architecture}\label{sec:2.4}

For the $(f_{7/2})^4$ space, the shell model eigenvalues are linear combinations of two-body interaction matrix elements, as well as the softmax input in the NN model. Therefore, the calculation process of the ground state spin determination is similar in both models. No need to include a hidden layer is required, as mentioned in Section \ref{sec:sm}. Actually, the neural network model without a hidden layer already achieves up to 98\% accuracy in predicting ground-state spins in the $(f_{7/2})^4$ space.

For the $(h_{11/2})^4$ space, since certain eigenvalues display nonlinear correlations with the two-body interaction matrix elements, incorporating hidden layers into the model becomes essential in order to enhance prediction accuracy. We first added one hidden layer and empirically chose 64 as the number of hidden nodes for a test run. The results showed that its accuracy reached 97\%, which is a satisfactory outcome.

For the remaining four spaces, the arbitrary accuracies are not always optimal. Therefore, we made multiple attempts to improve the prediction accuracy of our neural network classification model by adding more hidden layers and increasing the number of neural nodes in the $^{18}$Ne, $^{20}$Ne, $^{22}$Ne, and $^{46}$Ca model spaces.

\begin{figure}[!htb]
	\includegraphics
	[width=1.0\hsize]
	{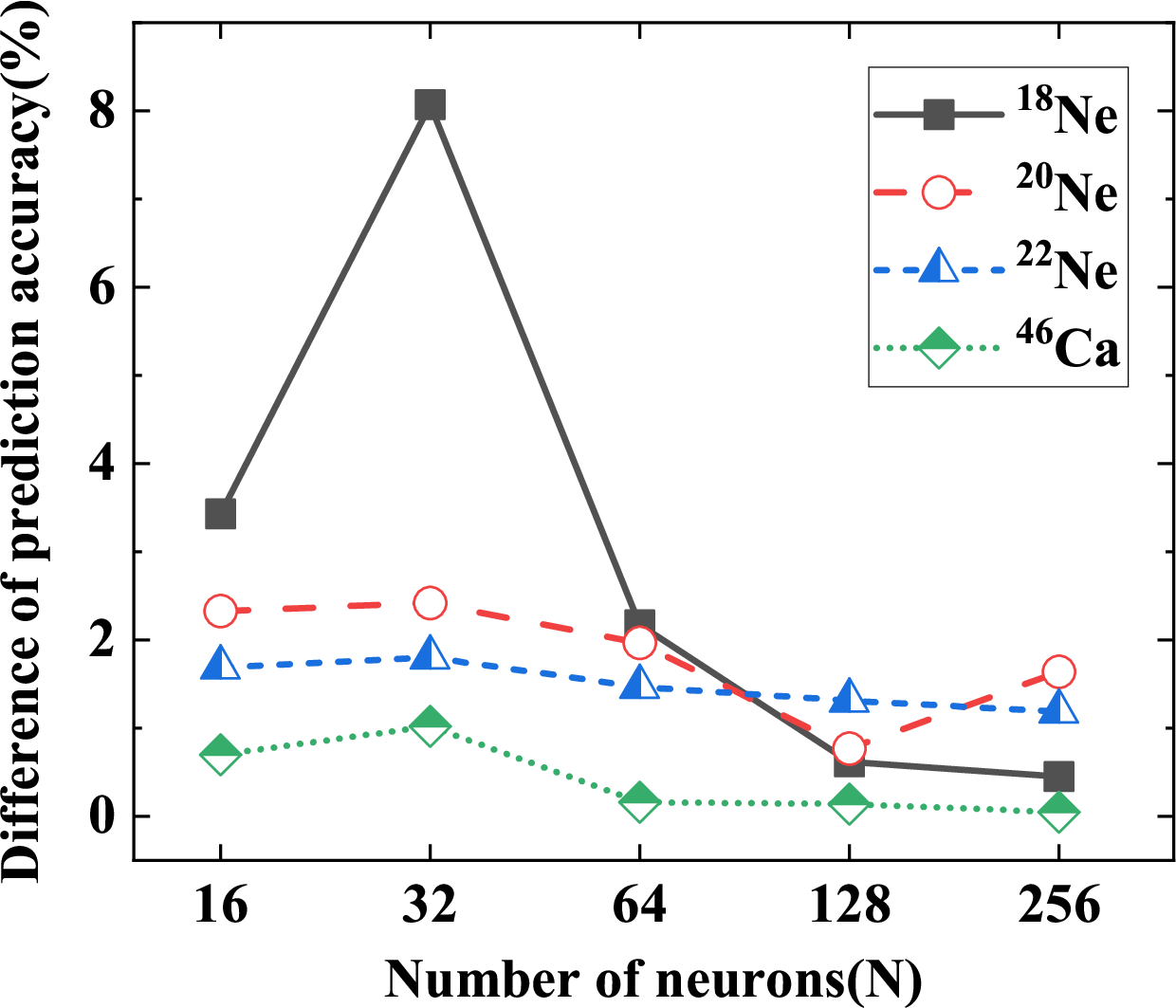}
	\caption{(Color online) Difference in prediction accuracy between models employing $N/2$ neural nodes and $N$ neural nodes for $^{18}$Ne, $^{20}$Ne, $^{22}$Ne, and $^{46}$Ca model spaces with a single hidden layer. 32 nodes are recommended.}
	\label{fig:2}
\end{figure}

Firstly, we observe the improvement in prediction accuracy when the number of neural nodes is doubled, indicated by the difference between the prediction accuracies with $N/2$ neural nodes and $N$ neural nodes, as depicted in Fig. \ref{fig:2}. The absence of negative differences in Fig. \ref{fig:2} suggests that doubling the number of neural nodes consistently results in improvement, as expected. It is further noted that the differences reach a peak when $N=32$ for all four model spaces. For $N> 32$, the prediction accuracy only demonstrates an improvement of $0\sim2$\%. Considering that more nodes entail additional computational overhead, we believe that 32 nodes may be the optimal and balanced choice for this study.

\begin{figure}[!htb]
	\includegraphics
	[width=1.0\hsize]
	{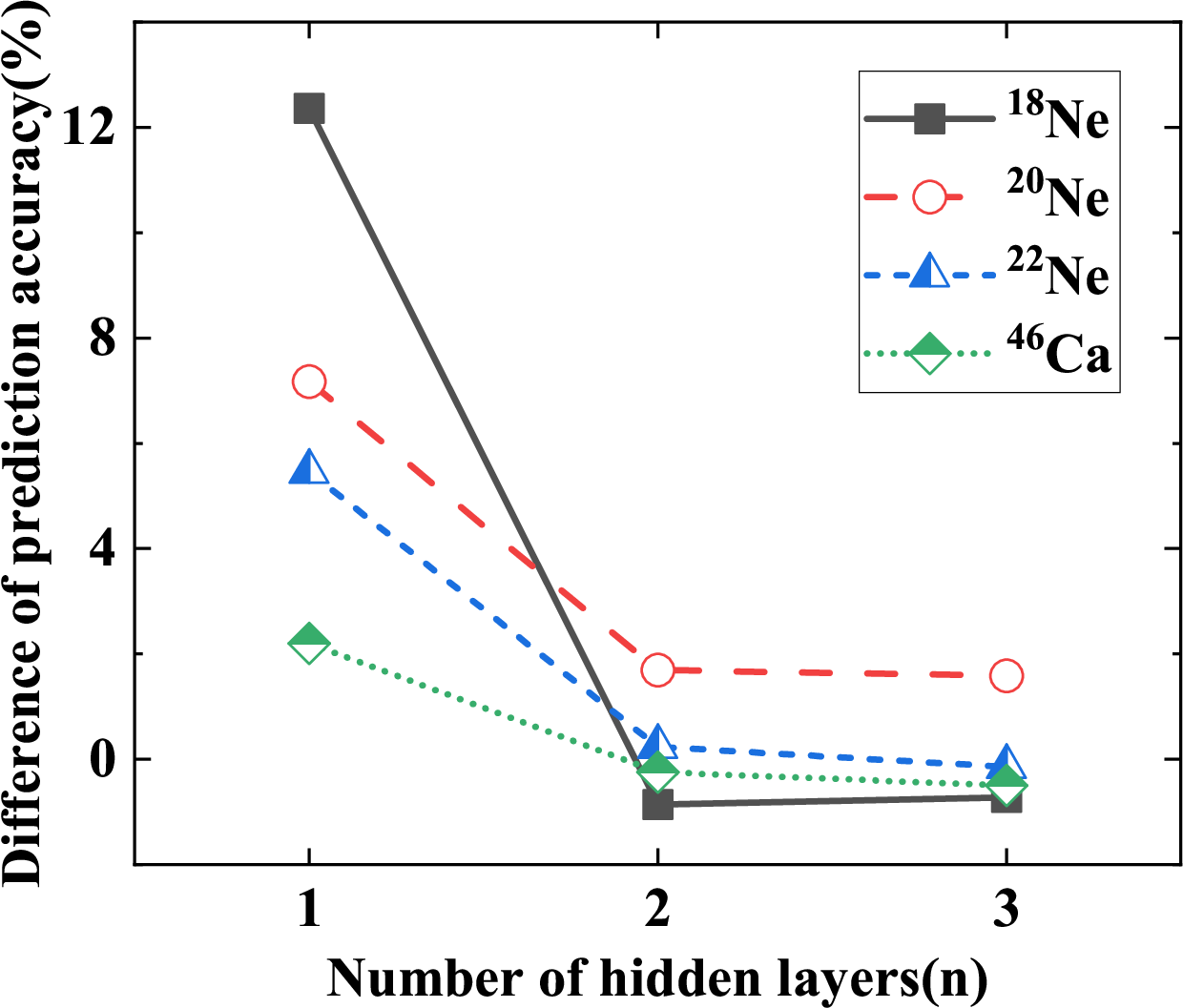}
	\caption{(Color online) Difference in prediction accuracy between networks employing $n$ hidden layers and networks with $n-1$ hidden layers for the $^{18}$Ne, $^{20}$Ne, $^{22}$Ne, and $^{46}$Ca model spaces, with 32 neural nodes in each hidden layer as recommended in Fig \ref{fig:2}. A single hidden layer is recommended.}
	\label{fig:3}
\end{figure}

Furthermore, we investigate the impact of hidden layers on prediction accuracy. By employing 32 neural nodes in each layer, as indicated in Fig \ref{fig:2}, we present the difference in prediction accuracy between networks with $n-1$ hidden layers and $n$ hidden layers against the layer number $n$ in Fig \ref{fig:3}, for the $^{18}$Ne, $^{20}$Ne, $^{22}$Ne, and $^{46}$Ca model spaces. A notable improvement in accuracy is observed with a single hidden layer, i.e., $n=1$. However, with the introduction of additional layers, such improvements diminish. As additional layers also consume computational resources, a single hidden layer can be an optimal choice.

The activation functions \cite{46} play a very crucial role in neural networks by learning the abstract features through nonlinear transformations. Common activation functions are Sigmoid (also called the Logistic function), Tanh (hyperbolic tangent) and ReLU functions, Table \ref {tab:2} presents the impact of different activation functions on the prediction accuracy of our neural network model. The model prediction accuracies of Tanh and ReLU function are very close for the five model Spaces. However, since the Tanh activation function includes the exponential operation, the computational overhead can be larger, we decide to use the ReLU function throughout the paper.

\begin{table}[!htb]
	\caption{Prediction accuracy (\%)  with three different activation functions as Sigmoid, Tanh, and ReLU. All the calculations are performed with a single 32-node hidden layer neural network model.}
	\label{tab:2}
	\begin{tabular*}{8cm} {@{\extracolsep{\fill} } cccccc}
		\toprule
		Activation function & $(h_{11/2})^4$ & $^{18}$Ne & $^{20}$Ne & $^{22}$Ne  & $^{46}$Ca \\ [0.5ex] 
		\midrule
		Sigmoid & 95.36 & 79.21 & 66.95 & 77.22 & 55.34 \\
		Tanh   &  96.15 & 85.10 & 67.69 & 78.39 & 55.67 \\
		ReLU   &  96.69 & 86.36 & 67.88 & 78.62 & 55.62 \\
		\bottomrule
	\end{tabular*}
\end{table}

In summary, the optimal network configuration for the $^{18}$Ne, $^{20}$Ne, $^{22}$Ne, and $^{46}$Ca model spaces consists of 1 hidden layer with 32 ReLU neural nodes. Our following analysis is all based on such a configuration.

\section{RESULTS AND ANALYSIS}

\subsection{Model comparison}

\begin{figure}[!htb]
	\includegraphics
	[width=1.0\hsize]
	{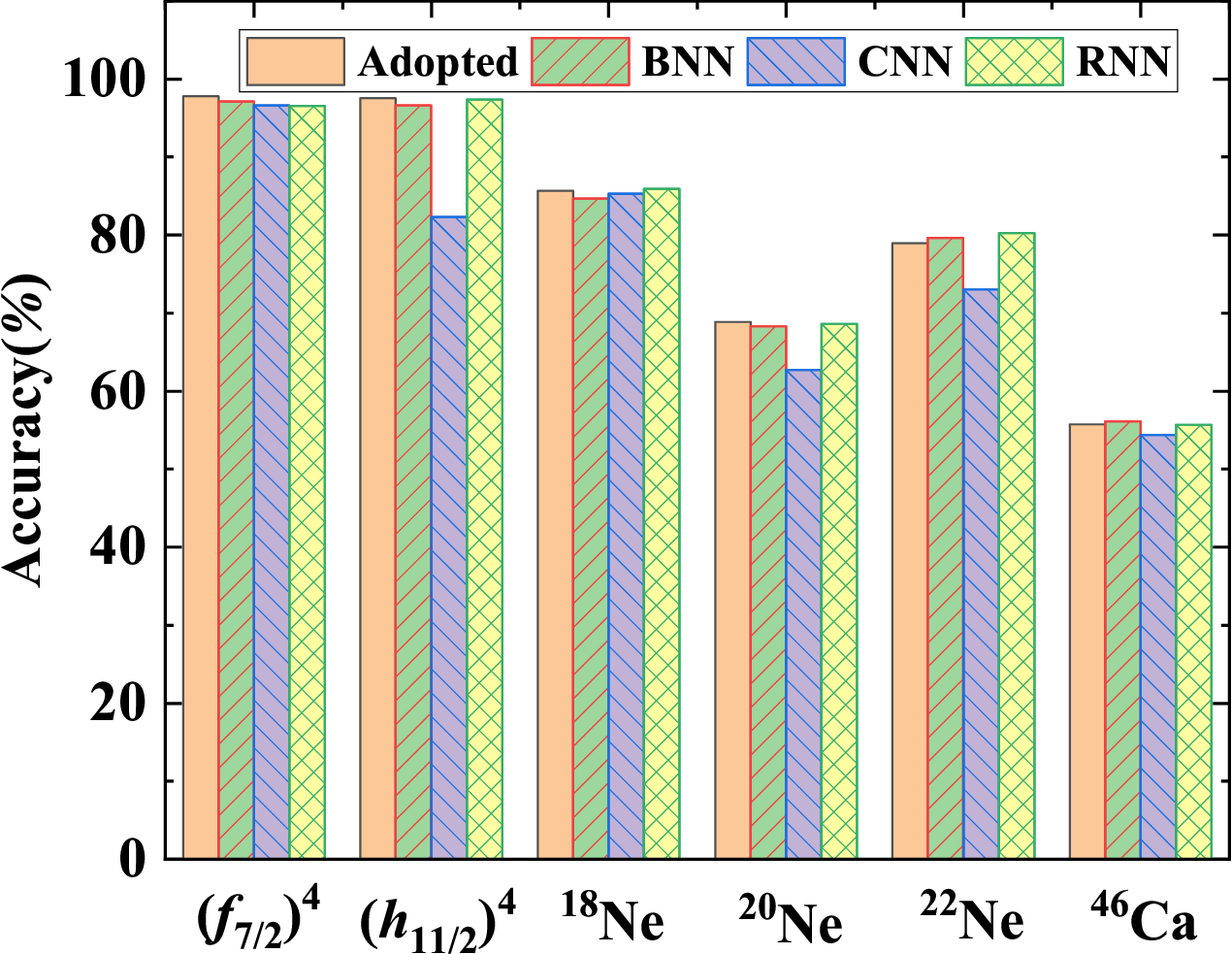}
	\caption{ (Color online) Prediction accuracies with different NN models. We adopt classic fully connected neural network as demonstrated in Fig. \ref{fig:1}. BNN stands for Bayesian neural network; CNN for convolutional neural network; RNN for recurrent neural network.}
	\label{fig:4}
\end{figure}

As demonstrated in Fig. \ref{fig:1}, we adopt a fully connected neural network model. However, given the recent application of Bayesian neural networks (BNN for short) on nuclear physics \cite{33,34,37,39}, as well as the great success of convolutional neural networks \cite{47,48,49,50} (CNN for short) and recurrent neural networks \cite{51,52,53} (RNN for short), we compare these 4 networks regarding the accuracy, as shown in Fig \ref{fig:4}.

The implementation of the BNN involves Bayesian sampling of weights and biases, facilitated by a variational inference algorithm to optimize model training. The model adopts 1000 iterations to update the loss and accuracy. In the prediction process, we sample 1000 times to yield more precise probability prediction results. The CNN includes a convolutional layer, a pooling layer, a fully connected layer, and a softmax layer. Especially, in the convolutional layer, the input channel is set as the number of input features, while the output channel is defined as 16, and the convolution kernel size is specified as 3 to facilitate feature extraction. Subsequently, in the pooling layer, the pooling kernel size and step size are set as 2 to reduce the dimension of the feature map. The fully connected layer maps the features extracted by the convolutional layer to the final classification result based on the task's feature dimension and category count. Throughout the construction process, the model's parameter settings were adjusted and optimized iteratively to ensure effective feature extraction and classification. A remarkable characteristic of the RNN is the architecture's ability to transmit and share information continuously through recurrent connections. Additionally, the network's calculation process involves defining the forward propagation function, which encompasses the output generated after input calculation and subsequent prediction through the fully connected layer and softmax function. All four models, including the adopted classic softmax model, share some consistent parameters, including a training-to-test set ratio of 2:1, a learning rate of 0.01, over 1000 epochs of training, the use of the Rectified Linear Unit (ReLU) activation function, and the Adam optimization function.
  
According to Fig. \ref{fig:4}, the CNN performs the worst, while both BNN and RNN exhibit similar accuracy to adopted network. However, the adopted networks have faster training speed, and requires the least computational resource. Therefore, we believe that the adopted network still is the optimal choice for our study.
  
\subsection{Feature selection}

Feature selection plays a crucial role in machine learning and data analysis, as it can enhance model performance, mitigate the risk of overfitting, boost computational efficiency, streamline model interpretation, and address issues related to noise and redundant information. It involves conducting correlation analysis to assess the relationship between each feature and the target variable. Subsequently, features exhibiting a strong correlation with the target variable will be selected, and others are excluded in further training. Given the nonlinear nature of both our feature data and label data, we utilized the Spearman correlation coefficient $\rho$ \cite{54} for feature selection as
\begin{equation}\label{eq6}
\rho=1-\frac{6 \sum d_i^2}{n\left(n^2-1\right)},
\end{equation}
where $d_i$ represents the difference of rank values of the $i$-th data pair, and $n$ represents the total number of observed samples.

We calculate the $\rho$ coefficients for four high-dimensional model Spaces, i.e., $^{18\sim 22}$Ne and $^{46}$Ca model spaces, and use different threshold sizes to select input features with strong correlation. Only feature with $\rho$ larger than threshold would be kept for further training. In Table \ref{tab:3}, we list the number of elements of the two-body matrix, i.e., the number of input features, over a certain threshold, and corresponding accuracy with such a threshold. 

\begin{table}[!htb]
	\caption{Model accuracies (\%) and input numbers under different feature selection thresholds. Threshold 0 means no feature selection.}
	\label{tab:3}
	\begin{tabular*}{8cm} {@{\extracolsep{\fill} } cccccccccccccccccccccccccccccccccccccccccccccc}
		\toprule
	\multicolumn{2}{c}{threshold}			&	0.1 	&	0.01   	&	0.001  	&	0  	\\
	\midrule
	\multirow{2}{*}{$^{18}$Ne}	&	accuracy	&	70	&	77	&	85	&	86	\\
	&	input number	&	5  	&	14 	&	28 	&	30 	\\
	\midrule
	\multirow{2}{*}{$^{20}$Ne}	&	accuracy	&	60	&	65	&	66	&	68	\\
	&	input number	&	6  	&	17 	&	25 	&	30 	\\
	\midrule
	\multirow{2}{*}{$^{22}$Ne}	&	accuracy	&	71	&	74	&	76	&	80	\\
	&	input number	&	4  	&	16 	&	25 	&	30 	\\
	\midrule
	\multirow{2}{*}{$^{46}$Ca}	&	accuracy	&	53	&	56	&	56	&	56	\\
	&	input number	&	1 	&	30 	&	80 	&	94 	\\
		\bottomrule
	\end{tabular*}
\end{table}

Based on Table \ref{tab:3}, as the threshold increases, the number of inputs after feature selection decreases as anticipated. However, this reduction in input number also corresponds to a decline in performance. Consequently, it is apparent that each inputted two-body matrix elements within the four model spaces have a profound impact on the output. Consequently, it is not recommended to exclude any of them in our network training.

\subsection{Accuracy}

\begin{figure}[!htb]
	\includegraphics
	[width=1.0\hsize]
	{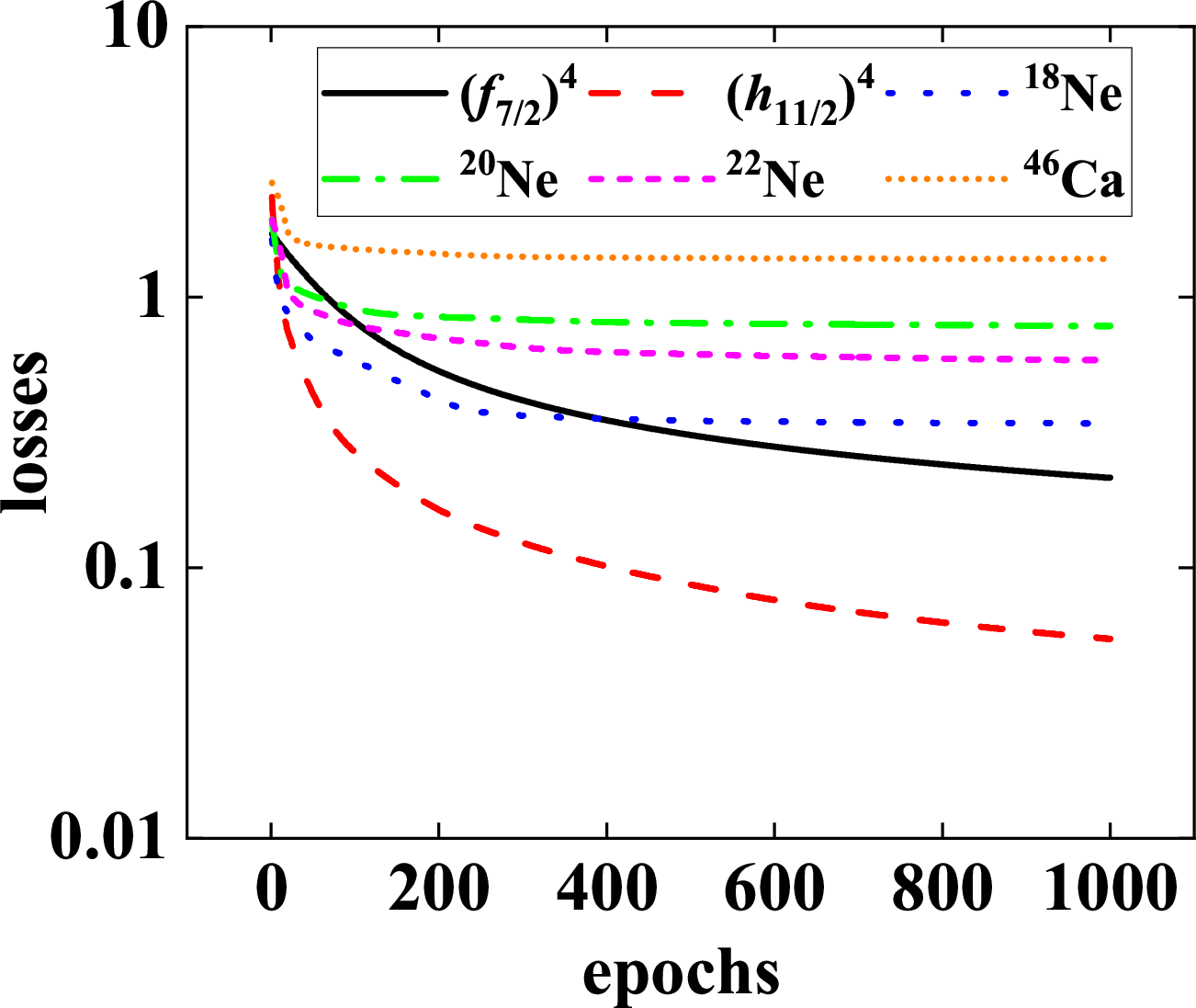}
	\caption{(Color online) Evolution of the loss functions during training.}
	\label{fig:5}
\end{figure}

Figure \ref{fig:5} present the evolution of the loss function during the training. As expected, the loss functions of the six model Spaces have converged, indicating that the network parameters is optimal. The loss values of two single-$j$ model spaces, i.e., $(f_{7/2})^4$ and $(f_{11/2})^4$, drop the most dramatically, which is not surprising, since single-$j$ spaces are more simpler than the rest of 4 models. We also note that $^{46}$Ca, $^{20}$Ne, $^{22}$Ne, $^{18}$N all converged with large loss value, corresponding to the unsatisfactory accuracy described in following Table \ref {tab:4}. Increasing the training epochs does not improve the accuracy.

Table \ref{tab:4} provides the correlation between the prediction accuracy (\%) of the neural network for ground-state spin and the dimension of the six model space under investigation. For the $(f_{7/2})^4$ space, since its shell model eigenvalues themselves are linear combinations of two-body interaction matrix elements, the neural network model is equivalent to linear regression and achieves a high prediction accuracy of up to 98\%. With one hidden layer, the accuracy reaches 97\% for the $(h_{11/2})^4$ space, although some eigenvalues in the $(h_{11/2})^4$ space exhibit nonlinear relationships with two-body interaction matrix elements.

\begin{table}[!htb]
	\caption{Model space dimensions, the prediction accuracy of the NN, and consistent rate of the $G-I$ correlations between the SM and the NN (see Subsection \ref{sec:gi} for definition).}
	\label{tab:4}
	\begin{tabular*}{8cm} {@{\extracolsep{\fill} } ccccccc}
		\toprule
		Model Space & $(f_{7/2})^4$ & $(h_{11/2})^4$ & $^{18}$Ne & $^{20}$Ne & $^{22}$Ne  & $^{46}$Ca \\ [0.5ex] 
		\midrule
		dimension      & 8 & 23 & 14 & 81 & 142 & 3952 \\
		accuracy (\%)   & 98 & 97 & 86 & 68 & 80 & 56 \\
		consistency (\%) & 100 & 100 & 100 & 60 & 80 & 74 \\
		\bottomrule
	\end{tabular*}
\end{table}

For the remaining four spaces, the accuracy significantly decreases as the dimensions increase. We obtain the Pearson correlation coefficient \cite{55} of -0.753 between the prediction accuracy (\%) and the dimension in logarithmic scale, indicating a negative correlation between the two variables. As the dimension of the space increases and the shell complexity grows, the neural network model's ability to predict ground-state spin diminishes. As shown in Fig. \ref{fig:2} and \ref{fig:3}, introducing more hidden layers or neural nodes does not significantly improve the performance of general classification neural networks. Thus, the generalization capability of the neural network is strongly challenged by the complexity of the quantum many-body system, and a more specialized neural network architecture and activation function should be designed according to the cfp coefficient property and diagonalization process, in order to accurately predict ground-state spin in the TBRE.

\begin{figure}[!htb]
	\includegraphics
	[width=1.0\hsize]
	{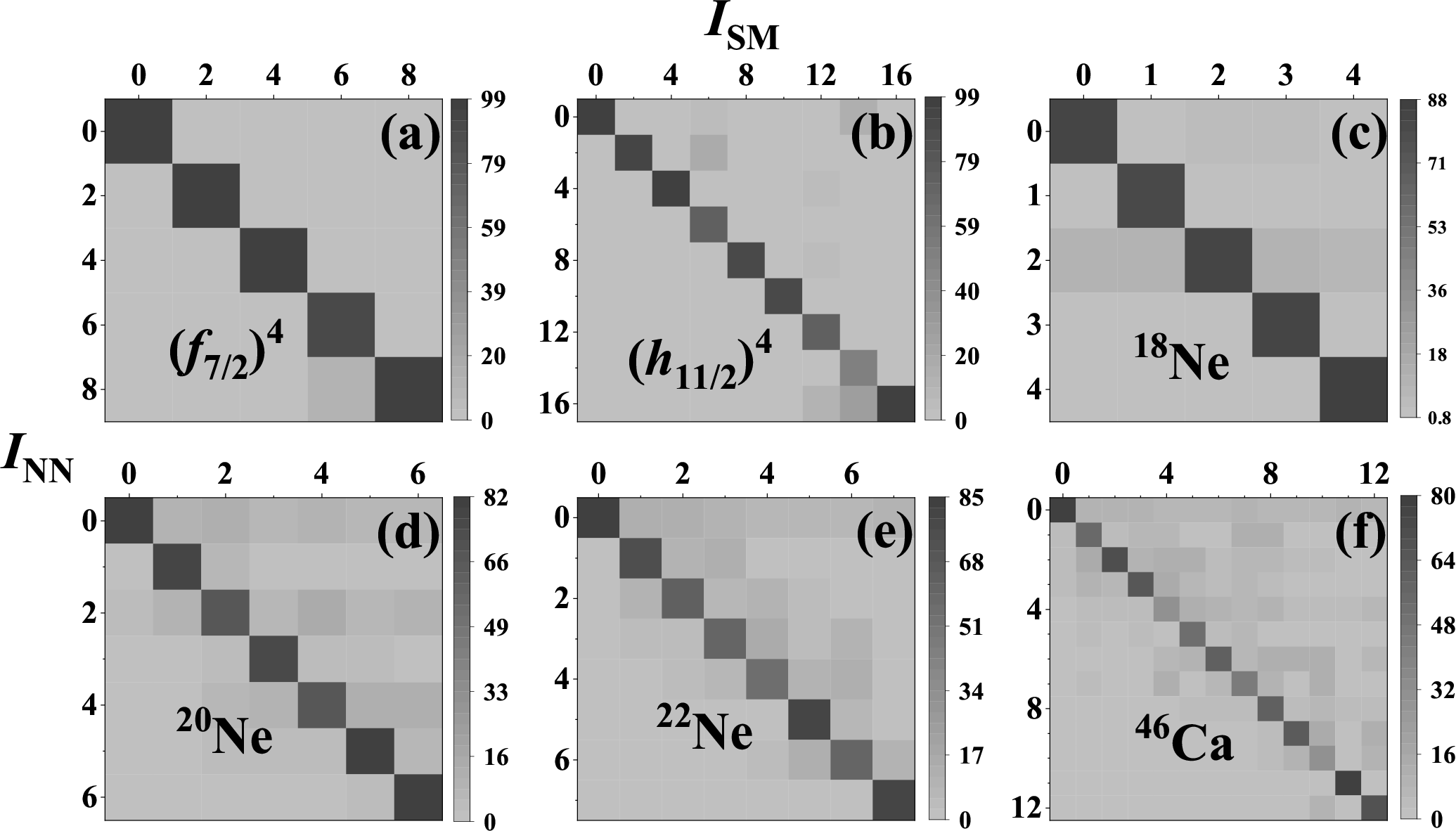}
	\caption{The confusion matrices for the prediction of ground-state spin using the neural network model in the $(f_{7/2})^4$, $(h_{11/2})^4$, $^{18}$Ne, $^{20}$Ne, $^{22}$Ne, and $^{46}$Ca TBRE calculations. The y-axis represents the ground-state spin predicted by the neural network ($I_{\rm NN}$), and the x-axis represents the ground-state spin obtained from the shell model calculations ($I_{\rm SM}$). The gray scale represent the probability of the shell model calculation yielding a ground-state spin of $I_{\rm SM}$ in the samples for which the neural network predicts a ground-state spin of $I_{\rm NN}$.}
	\label{fig:6}
\end{figure}

In order to gain a more detailed picture of the neural network model's prediction performance for TBRE samples with specific spin, Fig. \ref{fig:6} presents the confusion matrix for the neural network models of the six model spaces. In confusion matrices, the y-axis represents the ground-state spin predicted by the neural network ($I_{\rm NN}$), while the x-axis represents the ground-state spin obtained from the shell model calculations ($I_{\rm SM}$). The gray scale indicates the probability of the shell model calculation yielding a ground-state spin of $I_{\rm SM}$ in the samples, for which the neural network predicts a ground-state spin of $I_{\rm NN}$. The main diagonal of the confusion matrix appears predominantly dark, indicating a reasonably high degree of consistency between the neural network and the shell model for specific ground-state spin. From a statistical perspective, the neural network has captured some correlation between the ground-state spin and two-body interaction matrix elements of the TBRE.

Furthermore, based on the data from Table \ref{tab:4}, we notice that the prediction accuracy for the ground-state spin of the $^{20}$Ne nucleus is lower than that of the higher-dimensional $^{22}$Ne. This finding aligns with the observations in Fig. \ref{fig:6}. Specifically, it can be seen that for $^{20}$Ne, the difference in colors between the main diagonal and other regions is less pronounced than in other nuclei. This suggests that the prediction of ground-state spin in $^{20}$Ne space poses greater challenges to the neural network, which may be related to some special property of the $^{20}$Ne cfp coefficients. Further exploration of the specific multi-body complexity features in $^{20}$Ne space is desirable.

\begin{figure}[!htb]
	\includegraphics
	[width=1\hsize]
	{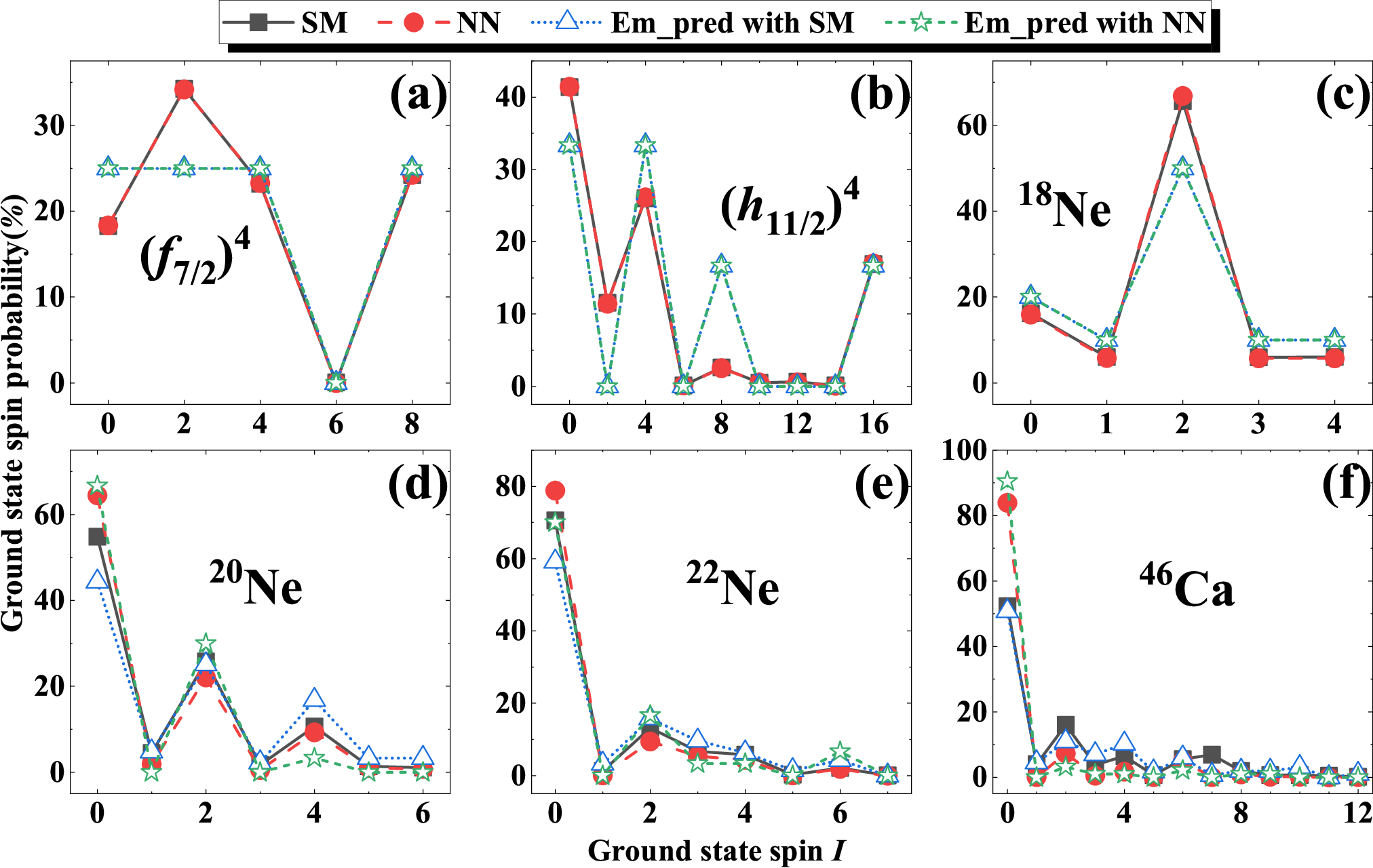}
	\caption{(Color online) Distribution of the ground-state spin $I$ ($P_I$) for $(f_{7/2})^4$, $(h_{11/2})^4$, $^{18}$Ne, $^{20}$Ne, $^{22}$Ne, and $^{46}$Ca. The black square represents $P_I$ from shell model calculations with random interactions. The red circle represents $P_I$ predicted by the neural network model. The blue triangle represents $P_I$ obtained with the empirical approach \cite{28} was applied to the shell model. The olive star represents $P_I$ obtained with the empirical approach was applied to the neural network model.}
	\label{fig:7}
\end{figure}

To further evaluate the statistical performance of the neural network model, Fig. \ref{fig:7} presents the distribution of ground-spin spins $ I$ ($P_I$) using both the shell model and the well-trained neural network model with random interactions. The neural network model shows good agreement with the shell model in all model spaces. The neural network has partially succeeded in capturing the robust statistical properties of the TBRE.

\subsection{$G-I$ correlation}\label{sec:gi}

To predict $P_I$ in TBRE, Zhao et al. proposed a general empirical approach \cite{19}. Their approach involves setting one of the two-body interaction matrix elements to -1 and the rest to 0. Such determined interaction is then inputted into the shell model, and the output ground-state spin $I$ is recorded. If there are $N$ independent two-body interaction matrix elements in the model space, the process is repeated $N$ times. Each time has a different matrix element equal to -1. Finally, the number of times the spin $I$ is observed as the ground state spin in the $N$ numerical experiments is represented as $N_{I}$. The probability of a spin $ I$ being in the ground state can then be estimated as follows:
\begin{equation}\label{eq6}
	P_I=N_{I} / N.	
\end{equation}
{ The empirical approach \cite{19} attributes the "specific spin $I$ as the ground-state spin" to a few two-body interaction matrix elements.} If there are relatively more two-body interaction matrix elements responsible for the spin $ I=0$, then the empirical rule provides a phenomenological explanation for the dominance of the ground state with zero spin. 

{ We note that  the empirical approach hints the correlation between two-body interaction matrix elements and ground-state spin, and it's the correlation decides the ground-state spin distribution, as shown in Fig. \ref{fig:7}.} Thus, the NN model with good prediction of the TBRE ground-state spin distribution should also produce similar correlation between two-body interaction matrix elements ($G^{JT}_{j_1j_2;j_3j_4}$) and ground-state spin ($I$) to the shell model. Therefore, we need to compare such element-spin ($G-I$) correlations in the shell model and those in the NN model.

\begin{table}[!htb]
	\caption{Ground state spin($I$) from the shell model and the NN model in $(f_{7/2})^4$ and $(h_{11/2})^4$ model spaces, with inputted $G^J=-1$ for some specific $J$, and other $G^J$s equal to 0, where  $G^J$ denotes the two-body interaction matrix element $G^J_{jj;jj}$, as defined in Eq. (\ref{eq1}). This table presents the correlation between the two-body interaction matrix elements and the ground-state spin in the empirical approach.}
	\label{tab:5}
	\begin{tabular*}{8cm} {@{\extracolsep{\fill} } ccrcc}
		\toprule
		&\multicolumn{2}{c}{$(f_{7/2})^4$}&\multicolumn{2}{c}{$(h_{11/2})^4$} \\
		\cmidrule(r){2-3}\cmidrule(r){4-5}
		$G_J$ & SM & NN & SM & NN\\
		\midrule
		$G^{0}$   & 0 & 0 & 0 & 0 \\
		$G^{2}$   & 4 & 4 & 4 & 4 \\
		$G^{4}$   & 2 & 2 & 0 & 0 \\
		$G^{6}$   & 8 & 8 & 4 & 4 \\
		$G^{8}$   &   &   & 8 & 8 \\
		$G^{10}$  &   &   & 16 & 16 \\
		\bottomrule
	\end{tabular*}
\end{table}

\begin{table}[!htb]
	\caption{Same as Table \ref{tab:5}, except for $^{18}$Ne, $^{20}$Ne, $^{22}$Ne, with $G^I_{j_1j_2;j_3j_4}$ as the matrix elements of the two-body interaction, where the subscription $j_1,~j_2,~j_3,~j_4$ equal 1, 2, 3, corresponding to $s_{1/2}$, $d_{3/2}$, and $d_{5/2}$ orbits in $sd$ shell, respectively. $I=0 \sim 4$ in this table represents the degenerate states with spin 0, 1, 2, 3, and 4 from the shell model. The inconsistency between the neural network model and shell model is highlighted in bold.}
	\label{tab:6}
	\begin{tabular*}{8cm} {@{\extracolsep{\fill} } ccrccccc}
		\toprule
		&\multicolumn{2}{c}{$^{18}$Ne}&\multicolumn{2}{c}{$^{20}$Ne}&\multicolumn{2}{c}{$^{22}$Ne} \\
		\cmidrule(r){2-3}\cmidrule(r){4-5}\cmidrule(r){6-7}
		$G^J_{j_1j_2;j_3j_4}$ & SM & NN & SM & NN & SM & NN\\
		\midrule
		$G^{0}_{1111}$ &0 & 0   & 0$\sim$4 & 0  & 0$\sim$6 & 0\\
		$G^{0}_{1122}$ &0 & 0   & 0,2,4    & 0  & 0,2,4 & 0\\
		$G^{0}_{1133}$ &0 & 0   & 0        & 0  & 0,2 & 0\\
		$G^{0}_{2222}$ &0 & 0   &0,2$\sim$4& 0  &0$\sim$5  & 0\\
		$G^{0}_{2233}$ &0 & 0   & 0 & 0         & 0 & 0\\
		$G^{0}_{3333}$ &0 & 0   & 0 & 0         & 0$\sim$2 & 0\\
		$G^{1}_{1212}$ &1 & 1   & 1 &\pmb{0}& 0 & 0\\
		$G^{1}_{1223}$ &1 & 1   & 2 &\pmb{0}& 0 & 0\\
		$G^{1}_{2323}$ &1 & 1   & 0 & 0         & 0 & \pmb{3}\\
		$G^{2}_{1212}$ &2 & 2   & 0,2 & 0       & 0 & 0\\
		$G^{2}_{1213}$ &2 & 2   & 2 &\pmb{0}& 2 & 2\\
		$G^{2}_{1222}$ &2 & 2   & 1$\sim$4  &\pmb{0}& 0$\sim$6 & 0\\
		$G^{2}_{1223}$ &2 & 2   & 0 & 0         & 0 & \pmb{2}\\
		$G^{2}_{1233}$ &2 & 2   & 0 & 0         & 0 & 0\\
		$G^{2}_{1313}$ &2 & 2   & 4 &\pmb{2}& 0,2,4 & 2\\
		$G^{2}_{1322}$ &2 & 2   & 0 & 0         & 0 & 0\\
		$G^{2}_{1323}$ &2 & 2   & 0 &\pmb{2}& 0 & 0\\
		$G^{2}_{1333}$ &2 & 2   & 2 & 2         & 0$\sim$4 & 2\\
		$G^{2}_{2222}$ &2 & 2   & 0 & 0         & 0,2$\sim$4 & 0\\
		$G^{2}_{2223}$ &2 & 2   & 2 &\pmb{0}& 2,3 & \pmb{0}\\
		$G^{2}_{2233}$ &2 & 2   & 0 & 0         & 0 & 0\\
		$G^{2}_{2323}$ &2 & 2   & 2 &\pmb{0}& 0 & 0\\
		$G^{2}_{2333}$ &2 & 2   & 0 & 0         & 0 & 0\\
		$G^{2}_{3333}$ &2 & 2   & 2 &\pmb{0}& 0 & 0\\
		$G^{3}_{1313}$ &3 & 3   & 5 &\pmb{2}& 0,2,4 & \pmb{3}\\
		$G^{3}_{1323}$ &3 & 3   & 4 &\pmb{0}& 3 & \pmb{0}\\
		$G^{3}_{2323}$ &3 & 3   & 0 & 0          & 0 & 0\\
		$G^{4}_{2323}$ &4 & 4   & 6 & 6          & 6 & 6\\
		$G^{4}_{2333}$ &4 & 4   & 4 &\pmb{0}& 2,3 & \pmb{0}\\
		$G^{4}_{3333}$ &4 & 4   & 4 & 4         & 0 & 0\\
		\bottomrule
	\end{tabular*}
\end{table}

\begin{table}[!htb]
	\caption{Same as Table \ref{tab:6}, except for $^{46}$Ca. The subscription $j_1,~j_2,~j_3,~j_4$ equal 1, 2, 3, 4 corresponding to $p_{1/2}$, $p_{3/2}$, $f_{5/2}$, and $f_{7/2}$ orbits in $pf$ shell, respectively.}
	\label{tab:7}
	\begin{tabular*}{8cm} {@{\extracolsep{\fill} } cccccc}
		\toprule
		$G^J_{j_1j_2;j_3j_4}$ & SM & NN & $G^I_{j_1j_2;j_3j_4}$& SM & NN \\
		\midrule
		$G^{0}_{1111}$ &0$\sim$10  & 0    & $G^{2}_{3334}$ &0        & 0   \\
		$G^{0}_{1122}$ &0$\sim$10  & 0    & $G^{2}_{3344}$ &0          & 0 \\
		$G^{0}_{1133}$ &0,2$\sim$6 & 0    & $G^{2}_{3434}$ &4   &\pmb{0} \\
		$G^{0}_{1144}$ &0$\sim$4   & 0    &$G^{2}_{3444}$ &0          & 0 \\
		$G^{0}_{2222}$ &0$\sim$10  & 0    &$G^{2}_{4444}$ &2         & 2 \\  
		$G^{0}_{2233}$ &0,2,4,6    & 0   &$G^{3}_{1313}$&0,2,4&\pmb{3}\\
		$G^{0}_{2244}$ &0          & 0    & $G^{3}_{1314}$ &3&\pmb{0} \\
		$G^{0}_{3333}$ &0$\sim$6   & 0    & $G^{3}_{1323}$ &3&\pmb{0}\\
		$G^{0}_{3344}$ &0          & 0    & $G^{3}_{1324}$ &4&\pmb{0}\\
		$G^{0}_{4444}$ &0$\sim$4   & 0    & $G^{3}_{1334}$ &4&\pmb{0}\\
		$G^{1}_{1212}$ &0          & 0    & $G^{3}_{1414}$ &0,2,4$\sim$8&2\\
		$G^{1}_{1223}$ &0          & 0    & $G^{3}_{1423}$ &0  & 0\\
		$G^{1}_{1234}$ &0,9        & 0    & $G^{3}_{1424}$ &0  & 0 \\
		$G^{1}_{2323}$ &0          & 0    & $G^{3}_{1434}$ &0  & 0\\
		$G^{1}_{2334}$ &0        & 0      & $G^{3}_{2323}$ &3 &\pmb{0}\\
		$G^{1}_{3434}$ &1,8&\pmb{0} & $G^{3}_{2324}$ &0 &\pmb{2}\\
		$G^{2}_{1212}$ &0          & 0    & $G^{3}_{2334}$ &0  & 0\\
		$G^{2}_{1213}$ &2&\pmb{0}    & $G^{3}_{2424}$ &0  & 0\\
		$G^{2}_{1222}$ &0$\sim$10  & 0    & $G^{3}_{2434}$ &0  & 0\\
		$G^{2}_{1223}$ &0,4,6      & 0    & $G^{3}_{3434}$ &0,10    & 0\\
		$G^{2}_{1224}$ &0          & 0 & $G^{4}_{1414}$ &0,2,4$\sim$6,8  &8\\
		$G^{2}_{1233}$ &0          & 0    & $G^{4}_{1423}$ &2 &\pmb{0}\\
		$G^{2}_{1234}$ &0,9        & 0    & $G^{4}_{1424}$ &6 &\pmb{0}\\
		$G^{2}_{1244}$ &0          & 0    & $G^{4}_{1433}$ &0    & 0\\
		$G^{2}_{1313}$ &0,2,4      &2    & $G^{4}_{1434}$ &1 &\pmb{0}\\
		$G^{2}_{1322}$ &0,2,4,6    & 0    & $G^{4}_{1444}$ &0$\sim$4 &0\\
		$G^{2}_{1323}$ &0          & 0     &$G^{4}_{2323}$&6 &\pmb{0}\\
		$G^{2}_{1324}$ &0          & 0    &$G^{4}_{2324}$  &2 &\pmb{0}\\
		$G^{2}_{1333}$ &0$\sim$8   & 0&$G^{4}_{2333}$&1$\sim$6&\pmb{0}\\
		$G^{2}_{1334}$ &2         &\pmb{0}& $G^{4}_{2334}$ &0,9    & 0\\
		$G^{2}_{1344}$ &2    &\pmb{0}& $G^{4}_{2344}$ &4 &\pmb{0}\\
		$G^{2}_{2222}$ &0$\sim$6   & 0    & $G^{4}_{2424}$ &0    & 0\\
		$G^{2}_{2223}$ &1,2,4,5&\pmb{0}    &$G^{4}_{2433}$&0  & 0\\
		$G^{2}_{2224}$ &0          & 0   &$G^{4}_{2434}$ &3  &\pmb{0}\\
		$G^{2}_{2233}$ &0,2$\sim$4,6&0    & $G^{4}_{2444}$ &0,2$\sim$4  & 0\\
		$G^{2}_{2234}$ &0          & 0    & $G^{4}_{3333}$ &0    & 0\\
		$G^{2}_{2244}$ &0,2$\sim$4 & 0    & $G^{4}_{3334}$ &0    & 0\\
		$G^{2}_{2323}$ &0          & 0    & $G^{4}_{3344}$ &0,10    & 0\\
		$G^{2}_{2324}$ &0          & 0    &$G^{4}_{3434}$  &0    & 0\\
		$G^{2}_{2333}$ &0,2$\sim$4,6& 0   & $G^{4}_{3444}$ &0    & 0\\
		$G^{2}_{2334}$ &0,10       & 0    & $G^{4}_{4444}$ &4    & 4\\
		$G^{2}_{2344}$ &0          & 0    &$G^{5}_{2424}$ &10 &\pmb{9}\\
		$G^{2}_{2424}$ &0,9        & 0    &$G^{5}_{2434}$ &0  & 0\\
		$G^{2}_{2433}$ &0          & 0    &$G^{5}_{3434}$ &1 &\pmb{0}\\
		$G^{2}_{2434}$ &0          & 0    &$G^{6}_{3434}$&12&\pmb{10}\\
		$G^{2}_{2444}$ &0          & 0    & $G^{6}_{3444}$ &0    & 0\\
		$G^{2}_{3333}$ &0          & 0    & $G^{6}_{4444}$ &6    & 6\\
		\bottomrule
	\end{tabular*}
\end{table}

{ In Tables \ref{tab:5}, \ref{tab:6} and \ref{tab:7}, the $G-I$ correlations between two-body interaction matrix elements ($G^J_{j_1j_2;j_3j_4}$ defined in Eq. (\ref{eq1})) and the ground-state spin ($I$), obtained from  the empirical approach applied to both the shell model and the NN model, are listed for the $(f_{7/2})^4$ and $(h_{11/2})^4$ model spaces, Ne isotopes, and $^{46}$Ca, respectively.}

According to Table \ref{tab:5}, in the $(f_{7/2})^4$ and $(h_{11/2})^4$ model spaces, the NN model produces perfectly consistent $G-I$ correlations with the shell model. This explained the agreement between the shell model and the NN model in Figures \ref{fig:6}(a,b) and \ref{fig:7}(a,b). In Table \ref{tab:6}, such perfect consistency could also be observed for $^{18}$Ne space, as a coordination with Figures \ref{fig:6}(c) and \ref{fig:7}(c). However, as the dimension increasing, the consistency for $^{20~22}$Ne in Table \ref{tab:6} and $^{46}$Ca in Table \ref{tab:7} gradually decreased. In $^{20}$Ne space, there are 12 inconsistent $G-I$ correlations out of 30 (40\%) between the SM and NN; in $^{22}$Ne 6 out of 30 (20\%); in $^{46}$Ca 24 out of 94 ($\sim$26\%). Such inconsistent rates are also correlated with the prediction accuracy for different model space, as shown in Table \ref{tab:4}.

{ Furthermore, the empirical approach is also applicable to the trained neural network model, by setting one of the inputs of the neural network to -1 and the rests to 0, and recording the ground-state spin ($I$) from the network.} Such an approach also reveals the correlation between interaction matrix elements and predicted ground-state spin, as well as the $P_I$ distribution, of the well-trained NN model. Table \ref{tab:5} presents the correlations between the matrix elements and the spin obtained from the shell model and the NN model. Correlations from both models are found to be identical, indicating that our neural network model has successfully learned the $G-I$ correlation as suggested by { the empirical approach. Thus, it can accurately reproduce the ground-state spin of the shell model for simple model.}

{ With the $G-I$ correlation from such a NN model, we make a count on the ground-spin $I$s emerging in the $G-I$ correlation, and then normalize them in to the $P_I$ distribution, as guided by the empirical approach. Such $P_I$ distributions based on the empirical approach with the NN model is depicted in Fig. \ref{fig:7}. It can be observed that the empirical approach from both the SM and NN models yields reasonably consistent $P_I$ distributions for all model spaces in this case.} This observation suggests that the neural network may effectively capture the correlation between two-body interaction matrix elements and the ground-state spin, which further explains its remarkable performance to reproduce the statistical property of the ground-state spin in the TBRE.

\section{CONCLUSION}

This study aims to utilize a neural network model to investigate the distribution of ground-state spin in the TBRE. By utilizing a Softmax classification neural network model, we try to reproduce the correlation between the matrix elements of the interaction and the ground-state spin, as labeled by the shell model, for the TBRE. The reliability of the NN model is analyzed based on its prediction accuracy and consistency with the empirical rule of the $P_I$ distribution.

Previous applications of neural network models in nuclear physics have primarily focused on their strong fitting capabilities. However, the analysis of the ground-state spin distribution in TBRE demonstrates the neural network's classification ability, which is rare in previous nuclear physics research. Furthermore, TBRE can provide extensive samples for training neural networks, potentially enhancing the performance of neural network model.

In our investigation, we pursued various strategies to enhance the network's performance, including introduction of BNN, CNN, and RNN, feature selection, and adjusting the number of neural nodes and hidden layers. However, none of these approaches could yield significant improvements with limited computational resources. Therefore, we must acknowledge that the quantum many-body problem remains a formidable challenge for neural network models. Addressing this challenge may necessitate further development of neural network architectures tailored for analyzing nuclear ground-state spin in the TBRE.
	
On the other hand, neural network models still offers some insights into specific robust statistical properties of ground-state spin. For instance, they can effectively capture the distribution of ground-state spin, as demonstrated in Fig. \ref{fig:7}. Moreover, the resulting confusion matrix exhibits dominant diagonal elements, indicating the consistency between the ground-state spin from the shell model and those predicted by the neural network model, as depicted in Fig. \ref{fig:6}. This success can be attributed to the neural network's capacity to replicate the correlation between ground-state spin and the two-body interaction matrix element in the shell model, as demonstrated in Table \ref{tab:5}, \ref{tab:6}, and \ref{tab:7}.


\end{document}